\documentclass[12pt]{iopart}

\usepackage{amsfonts}
\usepackage{amssymb}
\usepackage[dvips]{graphicx}

\usepackage{bbm}
 
\begin{document}
 
 
\title{Quantum memories and Landauer's principle}
 
\author{Robert Alicki}

\address{  
Institute of Theoretical Physics and Astrophysics, University
of Gda\'nsk,  Wita Stwosza 57, PL 80-952 Gda\'nsk, Poland}
\ead{fizra@univ.gda.pl}
\begin{abstract}
Two types of arguments concerning (im)possibility of constructing a scalable, exponentially stable
quantum memory equipped with Hamiltonian controls are discussed. The first type concerns ergodic properties of open Kitaev models which are considered as promising candidates for such memories. It is shown that, although the 4D Kitaev model provides stable qubit observables, the Hamiltonian control is not possible.  The thermodynamical approach leads to the new proposal of the revised version of Landauer's principle and suggests that the existence of quantum memory implies the existence of the perpetuum mobile of the second kind. Finally, a discussion of the stability property of information and its implications is presented.

\end{abstract}

\section{Introduction}
The most challenging idea in \emph{Quantum Information} is the possibility  of\emph{ fault-tolerant quantum information processing } which violates two rather fundamental principles. 
\par
The first one can be called \textbf{(Classical) Complexity--Theoretic Church--Turing Thesis }(CTT), which states (this formulation is not due to Church or Turing, but rather was  gradually  developed in complexity theory and "efficiently"  means up to polynomial-time reductions):\\
\emph{A probabilistic Turing machine can efficiently simulate any realistic model of computation.} 
\par
The second principle is \textbf{Bohr's Correspondence Principle}(BCP) expressed in a form:\\
\emph{ Classical physics and quantum physics give the same answer when the systems become large,}\\
or in other words:\\
\emph{For large systems the experimental data are consistent with classical probabilistic models.}
\par
Indeed, the Shor's algorithm, when realized in a fault-tolerant way, allows factoring of numbers in a polynomial time. This task  is believed to be unfeasible using classical probabilistic computers and hence CTT must be violated. On the other hand quantum computing should be scalable and therefore according to BCP at a certain scale a quantum computer should be described by a classical probabilistic model.

Although there exists a well-developed theory of fault-tolerant quantum computation (FTQC) based on \emph{error correction} schemes \cite{ZurekFT,DoritFT,Gottesman2000-FT-local,Preskill1998-rel}, its phenomenological assumptions are doubtful and in author's opinion not convincing \cite{AHHH2001,RAFNL,AlickiLZ2006-FT,Dya}. The basic idea of error correction is the following. One assumes that the (mixed) states of, say, single logical qubit can be identified with a subset $\mathcal{S}_q$ of the density matrices for a larger, but still controlled system and $T$ is a completely positive trace preserving map (CP-map) modeling errors (noise) caused by the interaction with an uncontrolled environment. The \emph{error correcting map} is another CP-map $R$ such that 
\begin{equation}
R T\rho_q =  \rho_q,\ \mathrm {for \ any}\ \rho_q \in \mathcal{S}_q .
\label{ra-ercor}
\end{equation}
As any CP-map $W$ is a contraction with respect to the trace norm i.e.
\begin{equation}
\|W\rho - W\rho'\|_1 \leq \|\rho - \rho'\|_1 , \ \mathrm{for \ any}\ \rho, \rho'
\label{ra-ercor1}
\end{equation}
we have
\begin{equation}
\|\rho_q - \rho_q'\|_1 =\|RT\rho_q - RT\rho_q'\|_1 \leq \|T\rho_q - T\rho_q'\|_1 \leq \|\rho_q - \rho_q'\|_1
\label{ra-ercor2}
\end{equation}
what implies $\|T\rho_q - T\rho_q'\|_1 = \|\rho_q - \rho_q'\|_1$ and hence the existence of a unitary operator $U$ such that
\begin{equation}
T\rho_q = U\rho_q U^{\dagger} \ \mathrm {for \ any}\ \rho_q \in \mathcal{S}_q .
\label{ra-ercor3}
\end{equation}
The equality (\ref{ra-ercor3}) means that the degrees of freedom which represent the logical qubit are not essentially affected by the noise and  therefore one should rather speak about \emph{error avoiding} than \emph{error correcting} schemes. The same argument is valid also for classical systems where  $U$ is a Koopman's unitary map induced by a measure preserving transformation on the phase-space. The above conclusion is fundamental for the feasibility of FTQC
which is equivalent to the existence of quantum systems - \emph{quantum memories} - which possess quantum subsystems sufficiently stable with respect to noise. 
\par
To be more precise, one can define a \emph{quantum memory} for a single \emph{encoded qubit} as the quantum system interacting with a heat bath at the temperature $T$ which consists of $N$ microscopic subsystems (e.g. spins-1/2, \emph{physical qubits}) such that a certain subalgebra of observables isomorphic to the algebra $M_2$ of a qubit is \emph{exponentially stable}   for the temperatures low enough. Exponential stability means that the relaxation times of the encoded qubit observables increase exponentially with $N$ and it is a necessary condition to  perform computations of an arbitrary, polynomial in $N$ length.
\par
A quantum memory admits \emph{Hamiltonian control} if one is able to implement the Hamiltonians of the form
\begin{equation}
H_{\mathrm{con}}(t)= X\otimes F_x (t)+ Y\otimes F_y (t)+ Z\otimes F_z (t) 
\label{ra-control}
\end{equation}
where $X, Y, Z$ are standard hermitian operators which span the algebra $M_2$ (analogs of Pauli matrices) and $F_x , F_y , F_z$ are time-dependent hermitian operators of a generic quantum system or, as a limit case, external classical fields.\\
\emph{ It seems that only an exponentially stable quantum memory with Hamiltonian control can be useful for fault-tolerant quantum information processing.}
\par
In the next Section the fundamental mechanism of stability for classical information is briefly discussed. Then, in the following Sections two approaches to the problem of existence of quantum memory are presented. The first, a constructive one, involves designing
of quantum $N$ spin-1/2 systems with particular Hamiltonians which are supposed to generate stable encoded qubit observables when weakly coupled to a heat bath at temperatures below a certain critical one. In particular, the results for 2D and 4D Kitaev models are discussed which show that while the later model contains exponentially stable qubit observables, their Hamiltonian control seems to be not feasible. The second approach is based on  thermodynamics. It is argued that the information processing based on stable information carriers leads to reexamination of the Landauer's principle. From this new formulation it follows that a quantum memory with a Hamiltonian control cannot be realized in Nature, because its existence implies the existence of  a \emph{perpetuum mobile of the second kind}. This raises questions about the very nature of the notion of information, both in the classical and quantum context.

\section{Stable classical memories}

The macroscopic world around us is full of bodies with well-defined positions and shapes and therefore able to encode information. Even much smaller objects, like macromolecules, exist in stable configurations which carry, for example, biologically relevant information. The mechanism which provides stability of such "memories" with respect to thermal noise is universal. Matter occupies rather local minima of free energy than a global minimum corresponding to a single uniform state. Those minima are separated by free energy barriers $F_N$ which are proportional to the systems sizes $F_N\sim N$. The Boltzmann factor $\exp{-F_N/kT}$ dominating the thermal transition rates between the minima yields life-times of such metastable states growing exponentially with $N$. Even for relatively small molecules these life-times became much longer than the time scale of cosmological processes in the Universe.
\par
As a simple example one can briefly discuss the classical Ising models which are able to encode  a single bit in terms of the magnetization sign.
Compare the mean-field Ising model ($\sigma_j=\pm 1$) with the Hamiltonian 
\begin{equation}
H_N^{mf}= -\frac{J}{2N}\sum_{i,j=1}^N \sigma_i\sigma_j 
\label{ra-Ising1}
\end{equation}
with the 1D-Ising Hamiltonian  
\begin{equation}
H_N^{1D}= -J\sum_{j=1}^N \sigma_j\sigma_{j+1} 
\label{ra-Ising2}
\end{equation}
The energy difference between two configurations
$+++++++++++++$ and $+++{\underbrace{----}_{k-times}}++++++$ is given by
\begin{equation}
\Delta E^{mf} = Jk +\frac{k^2}{2N}\ , \ \Delta E^{1D} = 2J
\label{ra-Ising3}
\end{equation}
and shows the mechanism of bit's protection against noise. While for the 1D Ising model the energy difference does not depend on the number $k$ of flipped spins, for the mean-field model it grows with $k$ (the same holds for 2D, 3D,...). At finite temperatures one has to take into account the entropy contribution to the free energy which suppress the effect of protection at high temperatures leading to the ferromagnetic phase transition phenomenon for 2D, 3D,.., and mean-field Ising models. The question arises: \emph{Does a similar mechanism can protect encoded qubit?}

\section{Kitaev models}

The Kitaev models in $D=2,3,4$ dimensions \cite{Kitaev:2003:2,Dennis:2002:4452K} are spin-1/2 models defined on a $D$-dimensional lattice with a toric topology. The Hamiltonian always possess the special structure:
\begin{equation}
H = -\sum_{s} X_s - \sum _{c} Z_c ,
\label{ra_KitH}
\end{equation}
where, $X_s = \otimes_{j\in s}\sigma^x_j$, $Z_c = \otimes_{j\in c}\sigma^z_j$ are products of Pauli matrices belonging to certain finite sets on the lattice, called \emph{stars} and \emph{cubes}. They are chosen in such a way that all $X_s , Z_c$ commute forming an abelian subalgebra $\mathcal{A}_{ab}$ in the total algebra of $2^N\times 2^N$ matrices. The noncommutative commutant of  $\mathcal{A}_{ab}$, denoted by $\mathcal{C}$, is  a natural candidate for  the subalgebra containing encoded qubit observables.
\par
The stability with respect to thermal noise has been studied within Markovian models with semigroup generators derived by means of the Davies weak coupling procedure. The obtained Markovian master equations possess all  properties necessary from the phenomenological point of view: any initial state relaxes to the Gibbs thermal equilibrium, detailed balance property holds \cite{Alicki:Lendi:2007}. For Kitaev models which are \emph{ultralocal} such derivations are mathematically sound  and simple.  This makes the analysis of spectral properties of the Davies generators feasible, but still too involved to be reproduced here; we refer the reader to \cite{Alicki:2007:6451,Alicki:2008:4584,AlickiH3-Kitaev} for details. 
\par
For the 2D Kitaev model the spectrum of the Hamiltonian (\ref{ra_KitH}) is particularly simple. The ground state  is four-fold degenerated and any excited state is fully characterized by one of the ground states and the positions on the lattice occupied by single excitations called \emph{anyons}. There are two types of anyons, corresponding to stars ($X$-type) or cubes ($Z$-type) and their total numbers are even. The relative energy of such a state with respect to ground states is equal to the total number  of  anyons.
\par
To describe thermal relaxation, one couples all spins to individual identical heat baths by Hamiltonian terms containing $\sigma^x_j, \sigma^z_j$. Then, the form of the Markovian master equation for the density matrix in the interaction picture, and with a uniform normalization of relaxation rates, is the following 
\begin{eqnarray}
&&\frac{d\rho}{dt}
 = \frac{1}{2} \sum_{j=1}^N \Bigl\{  \Bigl(
 [a_j, \rho\,a_j^\dagger] + [a_j\, \rho,a_j^\dagger]  \nonumber\\
 &&+ e^{-2\beta}\,( [a_j^\dagger, \rho\,a_j] + [a_j^\dagger\, \rho,a_j) \Bigr) 
 - [a_j^0, [a_j^0,\rho]] \Bigr\} \nonumber\\
&& + \frac{1}{2} \sum_{j=1}^N \Bigl\{  \Bigl(
 [b_j, \rho\,b_j^\dagger] + [b_j\, \rho,b_j^\dagger]  \nonumber\\
 &&+ e^{-2\beta}\,( [b_j^\dagger, \rho\,b_j] + [b_j^\dagger\, \rho,b_j]) \Bigr) 
 - [b_j^0, [b_j^0,\rho]]\Bigr\}.
\label{ra-mme}
\end{eqnarray}
Instead of defining here the operators $a_j, a_j^0, b_j, b_j^0$ their physical interpretation is outlined. The operator $a_j$ ($a_j^\dagger$) annihilates
(creates) a pair of type-$Z$ anyons attached to the site $j$ , while 
$a_j^0$ generates diffusion of anyons of the same type. Similarly, the operators $ b_j, b_j^\dagger , b_j^0$ correspond to the type-$X$ anyons.
\par
The structure of the Hamiltonian (\ref{ra_KitH}) and the master equation (\ref{ra-mme}) imply  that the $2D$-Kitaev model is equivalent to a gas of noninteracting particles which are created/annihilated in pairs and  diffuse. Heuristically, no mechanism of macroscopic free energy barrier between different phases
is present which could be used to protect even a classical information. Mathematically, it was proved that the dissipative part of the Davies generator (in the Heisenberg picture) possesses a spectral gap independent of the size $N$ and therefore no metastable observables exist in this system. 
\par
In contrast to the 2D case the 4D Kitaev model can be described by a picture similar to droplets in the 2D-Ising model  \cite{Dennis:2002:4452K}. The excitations of the system are represented by closed loops with energy proportional to the loops' length providing the mechanism
of a macroscopic energy barrier separating topologically nonequivalent spin configurations. The 3D model is an intermediate case,  this mechanism  works for the one type of excitations only. Therefore, only encoded "bit" is protected but not the "phase". The structure of the evolution equation, for all dimensions is always similar to (\ref{ra-mme}) with the operators $a_j^{\dagger}, b_j^{\dagger}$ creating excitations of two types and $a_j^0, b_j^0$ changing the shape of excitations but not their energy. 
\par
The rigorous arguments concerning 4D case are presented in the paper \cite{AlickiH3-Kitaev}. Generally, for all $D=2,3,4$ one can define \emph{bare qubit observables}  $X^{\mu}, Z^{\mu}\in\mathcal{C}$ where  $\mu = 1,...,D$. They are products of the corresponding Pauli matrices over topologically nontrivial loops (surfaces) which are not unique. However, only for 4D case there exist exponentially  metastable \emph{dressed qubit observables} ${\tilde X}^{\mu}, {\tilde Z}^{\mu}\in\mathcal{C}$ with  $\mu = 1,2,3,4$ related to the bare ones by the formulas
\begin{equation}
 {\tilde X}^{\mu} =X^{\mu}F^{\mu}_x ,\quad {\tilde Z}^{\mu} =Z^{\mu}F^{\mu}_z.
\label{ra-dressed}
\end{equation}
where $F^{\mu}_z , F^{\mu}_z$ are hermitian elements of the algebra $\mathcal{A}_{ab}$ with eigenvalues $\pm 1$. On the other hand, bare qubit observables are highly unstable with relaxation times $\sim \sqrt{N}$. The metastability of (\ref{ra-dressed})
is proved using the Peierls argument applied to classical "submodels" of the 4D-Kitaev model generated either by $-\sum_{s} X_s$ or $ - \sum _{c} Z_c $ and holds below certain critical temperature.
\par
Any metastable observable (say  ${\tilde X}^{\mu}$) is given by the following  procedure, which operationally determines its outcomes:\\
1. Perform a measurement of all observables $\sigma^x_j$.\\
2. Compute the value of  $X^{\mu}$ (multiply previous outcomes for spins belonging to the "surface" which defines $X^{\mu}$).\\
3. Perform a certain classical algorithm (polynomial in $N$)  which allows to compute from the $\sigma^x_j$- measurement data the value $\pm 1$ of "correction", i.e. the eigenvalue of
$F^{\mu}_x$, and multiply it by the bare value  to get the outcome of ${\tilde X}^{\mu}$.
\par
The procedure of above provides the structure of  4-qubit subalgebra  which is exponentially stable with respect to thermal noise below the critical temperature. However, the dressed qubit observables are given in terms of a certain algorithm involving  destructive and not repeatable measurements on individual spins. Those qubit observables cannot be used to construct Hamiltonians and therefore this type of memory is not equipped with Hamiltonian controls. In the author's opinion this fact reflects  fundamental difficulties with the implementation of the
idea of FTQC. This leads to a natural question: \emph{Can phenomenological thermodynamics provide restrictions or even no-go theorems for Quantum Memory, or generally for Quantum Information Processing?}\\

\section{Thermodynamics of information processing}

There exists a deep similarity between the complexity theory and thermodynamics. In both cases the predictions have asymptotic character with respect to the parameter $N$ which in the first case denotes the size of the input and in the second case the number of elementary physical constituents ( atoms, spins, photons,...), called \emph{particles}. To make these relations closer one needs to reformulate the laws of thermodynamics which are usually expressed in a natural language and have a common sense character.

In \cite{Alicki:2010}  the following reformulation has been proposed:\\
\textbf{Zero-th Law}:\\
\emph{Any $N$-particle system coupled to a thermal bath relaxes to the (possibly nonunique) thermodynamical equilibrium state at the bath's temperature with relaxation time growing at most polynomially in $N$.}\\
\textbf{Second Law}:\\
\emph{It is impossible to obtain an effective process such that the unique effect is the subtraction of a positive heat from a reservoir and the production of a positive work of the order of at least $k_B T$. By effective process we mean a process which takes at most polynomial time in the number of particles $N$.}\\
\textbf{Fluctuation Theorem}:\\
\emph{For a system consisting of $N$ particles the probability of observing during time $t$ an entropy production opposite to that dictated by the second law of thermodynamics decreases exponentially with $Nt$.}\\
\par
One should notice that the Zero-th Law provides now a proper definition of the thermal equilibrium which includes multiple thermodynamical phases and metastable configurations corresponding to local minima of the free energy.
Fluctuation Theorem \cite{Fluctuation} determines the limits of applicability of the phenomenological thermodynamics.

\subsection{Landauer's Principle}

The direct connection between thermodynamics and information theory is given by the Landauer's Principle\cite{Landauer} in the formulation of Bennett\cite{Bennett}:\\
\emph{Any logically irreversible manipulation of information, such as the erasure of a bit or the merging of two computation paths, must be accompanied by a corresponding entropy increase in non-information bearing degrees of freedom of the information processing apparatus or its environment}.\\
\emph{Specifically, each bit of lost information will lead to the release of an amount $k_BT\ln 2 $ of heat, where $k_B$ is the Boltzmann constant and $T$ is the absolute temperature of the circuit.} 
\par
There exist two kind of "proofs" of the Landauer's principle. The first, and correct one is essentially based on Szilard's discussion of the \emph{Maxwell demon}.  A Szilard engine consists of a single particle in a box coupled to a heat bath. Knowing  which half of the box is occupied by a particle one can close a piston unopposed into the empty half of the box, and then extract $k_BT\ln 2$ of useful work using isothermal expansion to its original equilibrium state. Assuming that the Second  Law holds the "measurement process" needs at least the same amount of work which is attributed by Bennett to a single bit erasure in the process of reseting of the measuring device.
\par
One should notice that the argument of above does not apply to the situation where a bit of information is encoded in two possible equilibrium (metastable) states. Namely,  after a measurement one cannot use the relaxation process, like isothermal expansion, to extract work and leave the system in a completely mixed state, because such relaxation needs exponentially long times.
\par
The other type of "proof", which is incorrect, employs the total entropy balance for the information bearing subsystem
plus the environment. In its simplest (and quantum) version the process of a bit's erasure is described by the map
\begin{equation}
\rho_{in}\otimes\omega_{in} \Longrightarrow  \rho_{out}\otimes\omega_{out},\ \rho_{in} =\frac{1}{2}(|0\rangle\langle 0| + |1\rangle\langle 1|),\ \rho_{out} =|0\rangle\langle 0| . 
\label{ra_biter}
\end{equation}
Here, the state $|0\rangle$ is a fixed reference state of the information bearing system and the initial unknown state
$\rho_{in}$ is completely mixed. The density matrices $\omega_{in}$ and $\omega_{out}$ describe the initial and final states of the heat bath. Assuming the validity of the Second Law for the total isolated system one obtains
the estimation for the entropy balance ($S(\rho) = -k_B \mathrm{Tr}\rho\ln\rho$)
\begin{equation}
S(\omega_{out} ) - S(\omega_{in}) \geq k_B\ln 2 
\label{ra_entropy}
\end{equation}
what implies that at least $k_B T\ln 2$ of heat is dissipated into the heat bath during the reseting process.
The weak point of this argument is related to "discontinuity" of entropy for large systems 
expressed in terms of Fannes inequality \cite{Fan} for two close density matrices of a system with $D$-dimensional Hilbert space and $\|\cdot\|_1$ denoting the trace norm
\begin{equation}
|S(\rho)- S(\rho')| \leq \|\rho -\rho'\|_1 \ln D - \|\rho -\rho'\|_1\ln(\|\rho -\rho'\|_1)\ .
\label{fan}
\end{equation} 
Due to (\ref{fan}) even an infinitesimally  small perturbation of the initial product state in (\ref{ra_biter}), which is always present, can cause a substantial change in the total entropy.
\par
Another important case illustrating this problem is the Markovian dynamics of an open system. Here the state of the total system is well-approximated by the product
$\rho(t)\otimes \omega_{B}$ where $\rho(t)$ is a solution of the Markovian master equation and $\omega_{B}$ is a fixed equilibrium state of a bath. Obviously, this product form is not consistent with the constant entropy of the total Hamiltonian system. The missing entropy is hidden in the small correction terms describing the residual system - bath correlations  and small local perturbations of the bath's state.
\par
Summarizing, one can trust only the argument proposed by Szilard which is entirely based on the Second Law and which leads to the following:
\par
{\bf Revised Landauer's Principle for Measurement }\\
\emph {A measurement which allows to distinguish between two different states of a system coupled to a heat bath:\\
a) needs at least $k_BT\ln 2$ of work if those states relax to their uniform statistical mixture,\\
b) does not need a net amount of work if those states are equilibrium  ones.}

\subsection{Quantum memory as perpetuum mobile}

Following \cite{Alicki:2010} one can construct a model of a \emph{perpetuum mobile of the second kind} under the assumption that
one possesses a single-qubit exponentially stable quantum memory with a Hamiltonian control. The device contains besides the memory a quantum version of the Szilard engine. It is a two-level system with controlled  energies of both levels. The system relaxes to the Gibbs state due to the interaction with a heat bath. Assume that the initial state is the one of those eigenstates and the initial energies are both equal to zero. If one knows which level is occupied then one can quickly increase the energy of the second level to the value $E>> k_B T$. In the next step one couples the system to a heat bath and slowly decreases the energy of the second level
back to zero. It is easily to compute that during this process an amount of work $W\simeq k_B T\ln 2$ is subtracted from the heat bath \cite{ALH32004}. In the standard setting one concludes that the cyclic process of acquiring information about the initial state needs at least the same amount of work. Here, the quantum memory is used to prepare a given initial state. Namely, according to the revised Landauer's principle a measurement of a state of  quantum memory in a given basis
does not cost work. Knowing the state of  memory one can swap it with the relaxing system and then start the process of extracting work. The swap operation is unitary and hence does not cost work and can be realized using the Heisenberg-type interaction Hamiltonian \cite{swap} between the memory and the relaxing qubit
\begin{equation}
H_{\mathrm{int}}(t)= \lambda(t)\bigl(X\otimes \sigma^x + Y\otimes \sigma^y + Z\otimes \sigma^z\bigr).  
\label{ra-heisenberg}
\end{equation}
Here $\sigma^k$ are Pauli matrices for the relaxing qubit and $\lambda(t)$ is a time-dependent coupling constant. The construction of such a Hamiltonian is guaranteed by the assumption of Hamiltonian control for the quantum memory. Finally, the net effect of the whole cyclic process is the subtraction of heat from the bath and the production of work what violates the Second Law of Thermodynamics.

\section{Two types of information?}

The arguments of the previous Section can be also applied to a classical stable memory. The revised Landauer's Principle is valid in the classical case also and it is not difficult to give an example of a classical Hamiltonian swap operation (it can be done for two classical systems with a single degree of freedom each and the Hamiltonian quadratic in position and momenta). To avoid the conflict with the Second Law one has to conclude that the Hamiltonian (reversible) gates are incompatible with stability of encoded information. Indeed, taking as a model of classical one-bit memory a particle in a double-well potential, it is rather obvious that friction is a necessary ingredient
to stabilize  information. Applying a Hamiltonian gate by kicking a particle from one well to the other one never reaches a final stable state but rather oscillations between two values of bits as long as friction is not at work. Obviously, in the existing computers all gates are strongly irreversible.
\par
It seems that stability of information is an important ingredient of its very definition. One often does not make 
difference between "pseudo-information" encoded e.g. in temporal positions of gas particles and  "information" encoded e.g. in a shape of a macroscopic body. The previous discussion on the Landauer's principle and the intuitive analysis of simple examples allow to characterize the fundamental features of both notions:\\
1)\textbf{ Pseudo-Information} is unstable with respect to  thermal noise; can be used to extract work from a heat bath and hence its acquiring costs work at least $k_B T\times (Shannon\ entropy)$; can be processed reversibly.\\
2)\textbf{ Information} is stable with respect to  thermal noise (at least below a certain temperature, with life-time scaling $\sim
\exp(\gamma N))$; its acquiring does not cost work ; its Shannon entropy has no thermodynamical meaning; must be processed irreversibly with a cost of a single gate $\sim \gamma Nk_BT$ of work.
\par
The intrinsic lack of stability with respect to thermal noise makes the practical applicability of pseudo-information very limited. In the author's opinion Quantum Information is, unfortunately, an example of pseudo-information.

\textbf{Acknowledgments}
The author thanks Micha\l\ Horodecki and  Hector Bombin for discussions. This work is supported by the  Polish Ministry of Science and Higher Education grant PB/2082/B/H03/2010/38.

\end{document}